\documentclass[twocolumn,pre,showpacs,floatfix]{revtex4}
\usepackage{graphicx}
\usepackage{amsfonts}
\usepackage{amssymb}
\usepackage{bm}

\newcommand{\be}{\begin{equation}}
\newcommand{\ee}{\end{equation}}
\newcommand{\bea}{\begin{eqnarray}}
\newcommand{\eea}{\end{eqnarray}}
\newcommand{\zncu}{ZnCu$_3$(OH)$_6$Cl$_2$}

\begin{document}

\title{Ground State of the Spin-1/2 Kagome Lattice Heisenberg Antiferromagnet}

\author{Rajiv R.~P.~Singh}
\affiliation{Department of Physics, University of California, Davis,
CA 95616, USA}

\author{David A.~Huse}
\affiliation{Department of Physics, Princeton University, Princeton,
NJ 08544, USA}
\date{\today}

% 75.10.Jm Quantized spin models

\begin{abstract}
Using series expansions around the dimer limit, we find that the ground
state of the spin-1/2 Heisenberg Antiferromagnet on the Kagome
Lattice appears to be a Valence Bond Crystal (VBC) with a 36 site
unit cell, and ground state energy per site $E=-0.433\pm0.001 J$. It
consists of a honeycomb lattice of `perfect hexagons'.  The energy
difference between the ground state and other ordered states with
the maximum number of `perfect hexagons', such as a stripe-ordered
state, is of order $0.001 J$. The expansion is also done for the 36
site system with periodic boundary conditions; its energy per site
is $0.005\pm 0.001 J$ lower than the infinite system, consistent
with Exact Diagonalization results.  Every unit cell of the VBC has
two singlet states whose degeneracy is not lifted to 6th order in
the expansion. We estimate this energy difference to be less than
$0.001 J$. The dimerization order parameter is found to be robust.
Two leading orders of perturbation theory give lowest triplet
excitations to be dispersionless and confined to the `perfect
hexagons'.
\end{abstract}

\maketitle

The spin-1/2 antiferromagnetic Kagome-Lattice Heisenberg Model
(KLHM) with Hamiltonian,
\begin{equation}
{\cal H}=J\sum_{\langle i,j\rangle} {\bf S}_i\cdot{\bf S}_j,
\end{equation}
is a highly frustrated quantum spin model \cite{review}. Its
properties have been studied by a wide variety of numerical and
analytical techniques
\cite{zeng90,singh92,leung93,elstner,lhuillier,waldtmann,mila,misguich05}.
Yet, the precise nature of the ground state remains a subject of
debate. Proposals have included a number of Valence Bond Crystals
(VBC) \cite{marston,nikolic,maleyev,auerbach} as well as spin-liquid
states with algebraic correlations \cite{hermele05,ran06}.  Recent
experimental work on the material \zncu  has attracted further
interest to this model \cite{helton07,ofer07,mendels07,imai07},
although this material is likely to also have significant
Dzyloshinski-Moria anisotropy. \cite{rigol07}

%%%%%%%%%%%%%%  FIGURE  %%%%%%%%%%%%%%%%%%%%%%%%%%%%%%%%%%%%%%%%%%%%%%
\begin{figure}[!htb]
\vspace{0.5cm}
\begin{center}
  \includegraphics[scale=.5]{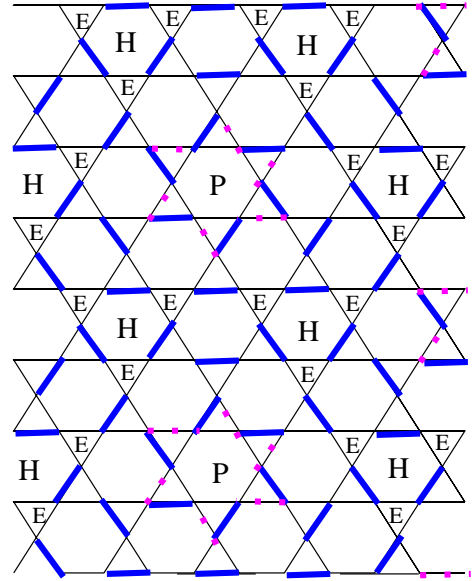}
\end{center}
\vspace{-0.2cm} \caption{\label{pattern} (Color online) Ground state ordering
pattern of low-energy (``strong'') bonds (blue) for the Kagome
Lattice Heisenberg Model. The perfect hexagons are denoted as H, the
empty triangles by E and the pinwheels as P. The two dimer coverings
of the pinwheels that remain degenerate to high orders of
perturbation theory are denoted by thick solid (blue) and dotted
(magneta) bonds.}
\end{figure}
%%%%%%%%%%%%%%%%%%%%%%%%%%%%%%%%%%%%%%%%%%%%%%%%%%%%%%%%%%%%%%%%%%%%%%%%

Here, we show that the ground state of KLHM appears to be a Valence
Bond Crystal with a 36 site unit cell. It consists of a Honeycomb
lattice of perfect hexagons as initially proposed by Marston and
Zeng \cite{marston}, discussed in more detail by Nikolic and Senthil
\cite{nikolic}, and shown in Fig~1.
%Because, the
%Kagome-Lattice is made out of corner sharing triangles, any dimer
%covering of the lattice is a low energy configuration.
In a dimer covering, all triangles that have a singlet valence bond
are locally in a ground state. As can be readily shown, any dimer
covering leaves one-fourth of all triangles in the Kagome lattice
empty.  All quantum fluctuations in the ground state originate from
these empty triangles, since it is only there that the singlet dimer
covering is not locally a ground state of the Hamiltonian.

We develop series expansions around an arbitrary dimer covering of
the infinite lattice using a Linked Cluster method \cite{book} and
compare the energies of various dimer coverings. To carry out the
expansions, all (``strong'') bonds that make up the dimer covering
are given an interaction strength $J$ and all other (``weak'') bonds
are given a strength $\lambda J$. Expansions are then carried out in
powers of $\lambda$ and extrapolated to $\lambda=1$ where all bonds
are equivalent in the Hamiltonian.  Following recent development of
the Numerical Linked Cluster scheme \cite{nlc}, we group together
all weak bonds belonging to each triangle. This significantly
simplifies the calculations:  %. The simplicity of the problem is such that
only 5 graphs contribute to the ground state energy to 5th order in
$\lambda$ (see Fig~2). The resulting series expansion for the ground
state energy shows surprisingly rapid convergence even at
$\lambda=1$.

%%%%%%%%%%%%%%  FIGURE  %%%%%%%%%%%%%%%%%%%%%%%%%%%%%%%%%%%%%%%%%%%%%%
\begin{figure}[!htb]
\vspace{0.5cm}
\begin{center}
  \includegraphics[scale=.4]{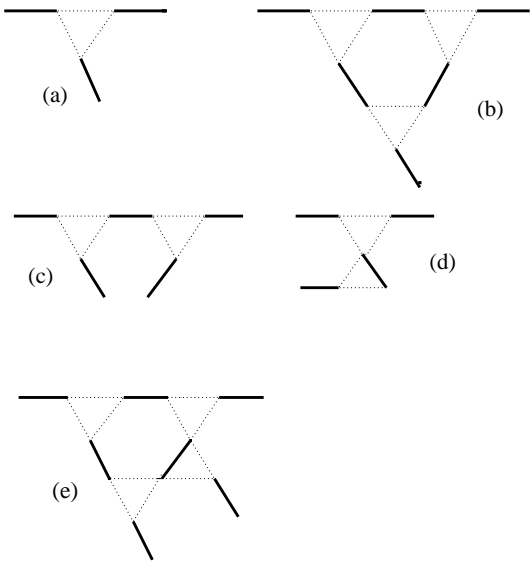}
\end{center}
\vspace{-0.2cm} \caption{\label{graphs} Topologies of the graphs
that contribute to the ground state energy to 5th order of
perturbation theory. Graphs (a), (b), (c), (d), and (e) begin
contributing in orders 2nd, 3rd, 4th, 4th, and 5th, respectively.}
\end{figure}
%%%%%%%%%%%%%%%%%%%%%%%%%%%%%%%%%%%%%%%%%%%%%%%%%%%%%%%%%%%%%%%%%%%%%%%%

To 2nd order in our expansion for the ground state energy, all dimer
coverings have the same energy.  At 0th order, the singlet dimers
each have energy $-0.75\ J$. These dimers can lower their energy
further by ``resonating'', but only on the empty triangles.  At 2nd
order, these pairwise dimer-dimer interactions lower the energy by
$0.28125\ J$ per empty triangle.  The degeneracy of all the dimer
configurations is finally lifted at third order, where there is
binding of 3 empty triangles into ``perfect hexagons''; the 3 dimers
around a perfect hexagon (see Fig. 2(b)) can then resonate together.
This process lowers the energy of each perfect hexagon at 3rd order
by $0.0703125\ J$.  At 4th order, resonances between two empty
triangles when they are connected by a single dimer bond as in Fig.
2(c) produce an additional binding energy of amount $0.01953125\ J$.

Thus at 4th order in our expansion we find that the ground state
comes from the dimer covering with the maximum number of perfect
hexagons, with neighboring hexagons arranged such that their empty
triangles share a dimer bond as much as possible. It follows that
once the positions and dimer coverings of two neighboring perfect
hexagons are picked, the rest uniquely fall on a Honeycomb lattice
of perfect hexagons. Furthermore, the dimerizations of all hexagons
are simple translations of one another. This leads to the Valence
Bond Crystal (VBC) arrangement shown in Fig.~1.  For an infinite
lattice, the ground state of this array of perfect hexagons is
24-fold degenerate, with 12 distinct translations, each of which has
2 reflections. As shown in Fig.~1, the arrangement into this VBC
also leaves a ``pinwheel'' of dimers in every unit cell. There are
two degenerate pinwheel dimer coverings, and this degeneracy is not
lifted to quite high orders of perturbation theory, implying that in
a lattice of $N$ sites there are at least $2^{N/36}$ states whose
energy difference, per site, is much less than $0.001\ J$.

We have carried the expansion to 5th order for the ground state
energy.  For the infinite lattice, the contribution at 5th order is
quite small (about $0.0005\ J$ per site) and slightly increases the
energy difference between the honeycomb and stripe VBC
\cite{nikolic} that was established at 4th order.

In addition to carrying out the expansions for the infinite system,
one can also study the ground state properties of finite systems
using the series expansion techniques.  The 36 site cluster with
periodic boundary conditions (PBC) studied by Leung and Elser
\cite{leung93} and Lecheminant {\it et al.} \cite{lhuillier}, does
accommodate the unit cell of the Honeycomb VBC but not that of the
stripe VBC.  For the honeycomb VBC dimer covering on this cluster,
the energy agrees with the infinite system through 3rd order.
However, at 4th order there are additional graphs that enter due to
the periodic boundary conditions, permitting 4 dimers to resonate
along a path through 4 empty triangles that winds once around the
periodic boundaries.  From comparing the series through 5th order
and using simple Pade estimates, we find the ground state energy for
the 36 site cluster to be lower than that of the infinite system by
$0.005\pm 0.001\ J$ per site, consistent with the Exact
Diagonalization result of an energy of $-0.438377\ J$ per site
\cite{leung93}.

The $\lambda=1$ energies for the infinite-lattice Honeycomb VBC, the
stripe VBC of Nikolic and Senthil \cite{nikolic}, and for the finite
36 site cluster with PBC and honeycomb VBC are listed in Table I,
summed through 5th order. Note the apparently rapid convergence of
the series, especially for the infinite lattice VBC states.

\begin{table}[h]
\caption{Ground state energy per site in units of $J$ for various
dimer states in perturbation theory.  At each order we sum the
contributions
through that order for $\lambda=1$.} %\vspace{0.5cm}
\begin{tabular}{rrrr}
\hline\hline
 Order &  Honeycomb VBC & Stripe VBC   & 36-site PBC   \\
   0   &    -0.375      &   -0.375     &   -0.375      \\
   1   &    -0.375      &   -0.375     &   -0.375      \\
   2   &    -0.421875   &   -0.421875  &   -0.421875   \\
   3   &    -0.42578125 &   -0.42578125&   -0.42578125 \\
   4   &    -0.431559245&   -0.43101671&   -0.43400065 \\
   5   &    -0.432088216&   -0.43153212&   -0.43624539 \\
\hline

\hline\hline
\end{tabular}
\end{table}

For the 36 site cluster, our considerations imply $48$ spin-singlet
states with very low energies per site within of order $0.001\ J$ of
the ground state, $24$ corresponding to the ground state degeneracy
of the thermodynamic system and $2$ each for the pin-wheel states.
But the other $48$ states with two `perfect hexagons' should also
fall below the lowest triplet state, whose excitation energy for the
36 site cluster translates to a per site value larger than $0.004\
J$. In the Exact Diagonalization studies about $180$ singlet states
are found below the lowest triplet. Since the binding energy for a
perfect hexagon translates into a per site energy of $0.002\ J$ for
the 36 site cluster, some singlet states with only one perfect
hexagon presumably can also have energies below the triplet gap in
that cluster.

We have also studied the local bond energies to 3rd order in
perturbation theory. To this order only four types of nearest
neighbor bonds, representing only 27 of the 72 bonds per unit cell,
have their expectation values changed from the 0th order values of
$-0.75\ J$ for strong bonds and $0$ for weak bonds.  These are: bond
type A, the strong bonds inside the perfect hexagons; bond type B,
the weak bonds inside the perfect hexagons; bond type C, the
weak-bonds in the empty triangles which do not form part of the
hexagon; and bond type D, the strong bonds that join two empty
triangles between perfect hexagons. Their expectation values are
listed in Table II.  The total ground state energy is about $-0.22\
J$ per bond.  Although the series for the total energy is converging
quickly to this value, the energies of individual bonds remain very
different, as is expected for a valence bond crystal.

\begin{table}[h]
\caption{ Expectation values $\langle S_i\cdot S_j\rangle$ for
various bond types}
\begin{tabular}{rrrr}
\hline\hline
 Bond &  0th order  &       2nd order          &    3rd order  \\
\hline
   A  &  -0.75      &       -0.5625            &   -0.515625   \\
   B  &   0         &       -0.1875            &   -0.257812   \\
   C  &   0         &       -0.1875            &   -0.1875     \\
   D  &  -0.75      &       -0.5625            &   -0.5625     \\
\hline\hline
\end{tabular}
\end{table}

Leung and Elser\cite{leung93} had also calculated the energy-energy
correlations in the ground state of the 36 site cluster. They had
noted that the correlations at largest distances qualitatively
matched the Valence Bond Crystal pattern after averaging over the
$48$ degenerate states. However, on a quantitative level there was
no correspondence with the VBC, as the dimerization pattern seen in
the Exact Diagonalization was considerably weaker. The modified
values of the bond energy expectation values, calculated by us, do
not significantly change this conclusion. We believe the reason for
the discrepancy is the following: In the 36 site cluster,
energy-energy correlations do not factorize even at the largest
distances. Due to the Periodic Boundary Conditions, even bonds which
are furthest away on the 36 site cluster can get correlated already
in 2nd order of perturbation theory. Thus it is not appropriate to
simply compare correlations $\langle (\vec S_i \cdot \vec S_j)(\vec
S_k \cdot \vec S_l)\rangle$ to the products $\langle\vec S_i \cdot
\vec S_j\rangle\langle\vec S_k \cdot \vec S_l\rangle$. In the
future, we hope to calculate and compare the energy-energy
correlations for the thermodynamic system and the 36 site cluster
within this dimer expansion.

We have also calculated two leading orders of perturbation theory
for the triplet excitation spectra. There are 18 elementary triplet
excitations per unit cell, corresponding to the 18 dimers. In 0th
order, these triplets are dispersionless and have excitation energy
$J$. To second order in perturbation theory, only 9 of the 18
triplets, consisting of the six belonging to the two perfect hexagons, and
the three that couple the perfect hexagons via empty triangles,
become dispersive and form a network on which excitations can move.
The other 9 triplets, consisting of the 6 pinwheel dimers and the 3
other dimers that do not touch empty triangles, can only perform
virtual hops at this order and thus remain dispersionless.  The
effective Hamiltonian for the nine dispersive states can be expressed in
terms of a $9\times 9$ matrix. Let $z_1=\exp{i \vec k \cdot \vec
r_1}$ with $\vec r_1=-4 \sqrt{3} \hat y$, and $z_2= \exp{i \vec k
\cdot \vec r_2}$ with $\vec r_2=-6\hat x -2\sqrt{3} \hat y$, and
$z_1^*$ and $z_2^*$ be their complex conjugates, with the lattice
oriented as in Fig. 1, with a nearest-neighbor spacing of unity.
Then, in second order perturbation theory, the triplet excitation
energies are the eigenvalues of the matrix
\begin{equation}
\Delta/J=(1-{5\over 8}\lambda^2)+({1\over 4}\lambda+{1\over
8}\lambda^2) M_1 +({1\over 32} \lambda^2) M_2,
\end{equation}
where the matrices $M_1$ and $M_2$ as a function of $z_1$ and $z_2$
are given in Table III and IV.

\begin{table}[h]
\caption{ Matrix $M_1$}
\begin{tabular}{rrrrrrrrr}
\hline\hline
  0&  -1&  -1&  -1&   1&   0&   0&   0&   0\\
 -1&   0&  -1&   1&   0&  -1&   0&   0&   0\\
 -1&  -1&   0&   0&  -1&   1&   0&   0&   0\\
 -1&   1&   0&   0&   0&   0&  -1&   0&   1\\
  1&   0&  -1&   0&   0&   0&   0&   $z_1$& $-z_1$\\
  0&  -1&   1&   0&   0&   0&   $z_2$& $-z_2$&  0\\
  0&   0&   0&  -1&   0&   $z_2^*$&  0&  -1&  -1\\
  0&   0&   0&   0&   $z_1^*$& $-z_2^*$& -1&  0&  -1\\
  0&   0&   0&   1&   $-z_1^*$& 0&   -1&  -1&  0\\
\hline\hline
\end{tabular}
\end{table}

Note that this implies that two triplets have the lowest energy and
they are each confined to one of the two perfect hexagons and thus
remain dispersionless. The triplet gap becomes
$1-(1/2)\lambda-(7/8)\lambda^2$, which adds up to $-3/8$ at
$\lambda=1$ if we truncate at this order. This can be understood as
arising from two factors: The hexagon is like a one-dimensional
Alternating Heisenberg Chain and in that case the gap is known to be
$1-(1/2)\lambda-(3/8)\lambda^2$ \cite{AHC}.  There is a slight
difference here with respect to the Alternating Heisenberg Chain
because of the additional neighbors. The diagonal term is doubled
but the second neighbor hop is absent, so that the overall result
for the gap is the same. In addition, each strong bond has its two
ends both connected to the same end of another dimer to which a
triplet on that strong bond may make a virtual hop.  This is like
the triplets in the Shastry-Sutherland model \cite{SS1,SS2}.  Due to
this, and the fact that no such process happens in the ground state,
there is a dispersionless reduction of the spin gap of magnitude
$-(1/2)\lambda^2$. At 2nd order these two types of contributions
simply add.  Thus we find that although the series for the ground
state energy shows apparently strong convergence, the series for the
spin gap does not, although the latter statement is about only the
first two orders in the expansion.  Further study of the triplet
excitations is left for future work.

Our results have important implications for the finite temperature
properties of the model. First of all, since the difference in
energy of the Honeycomb VBC state and other dimer states (such as
stripes) is less than $0.001 J$ per site, any finite temperature
transition in to a Honeycomb VBC phase should occur at a temperature
of this order or lower. Secondly, given the large number ($24$) of
degenerate ordering patterns, the phase transition seems likely to
be first order. Third, since the attraction between the empty
triangles is only of order $0.02 J$, above $T\approx 0.02 J$, all
dimer configurations should have comparable Boltzmann weight, giving
rise to a dimer liquid regime at intermediate temperatures. Fourth,
the specific heat and entropy of the KLHM should have structure down
to $T/J<0.001$. It was found in the high temperature expansion study
\cite{elstner} that a naive extrapolation of the high temperature
series down to $T=0$ led to a finite ground state entropy.
Furthermore, several studies have suggested multiple peaks in the
specific heat \cite{elser, misguich05}.  Our work shows at least
$2^{N/36}$ very low energy states of the `pinwheels', for an
$N$-site system, and many other very low lying states, giving
further support to these results.

\begin{acknowledgments}

This work was supported by the US National Science Foundation,
Grants No.\ DMR-0240918 (R. S.) and DMR-0213706 (D. H.) and
PHY05-51164.

\end{acknowledgments}

\begin{widetext}

\begin{table}[h]
\caption{ Matrix $M_2$}
\begin{tabular}{rrrrrrrrr}
\hline\hline
  0 &   0 &   0  &   1 &  -1 &   0  &   -1  &   $-z_2$  &  $1+z_2$ \\
  0 &   0 &   0  &  -1 &   0 &   1  &  $1+z_1$ &   $-z_1$  &  -1   \\
  0 &   0 &   0  &  0  &   1 &  -1  &   $-z_1$ &   $z_1+z_2$&  $-z_2$  \\
  1 &  -1 &   0  &  0  &  $1+z_2^*$&$1+z_1^*$ &  1  &    0   &  -1   \\
 -1 &   0 &   1  &  $1+z_2$&  0 &$1+z_1^*z_2$& 0  &  $-z_2$  &   $z_2$  \\
  0 &   1 &  -1  &  $1+z_1$&$1+z_1z_2^*$& 0    &$-z_1$ &    $z_1$  &   0   \\
 -1 &  $1+z_1^*$&$-z_1^*$&   1  & 0   & $-z_1^*$   & 0  &    0   &   0   \\
 $-z_2^*$& $-z_1^*$ &$z_1^*+z_2^*$&0  &$-z_2^*$ &  $z_1^*$   & 0  &    0   &   0   \\
 $1+z_2^*$& -1  &$-z_2^*$ &  -1 & $z_2^*$ &  0     & 0  &    0   &   0   \\
\hline\hline
\end{tabular}
\end{table}
\end{widetext}

\end{document}